\documentclass[
preprint,
amsmath,amssymb,
aps,
]{revtex4-1}

\usepackage{graphicx}
\usepackage{multirow}
\usepackage{bm}
\usepackage{float}
\usepackage{color}
\usepackage{textcomp}
\usepackage{hyperref}

\usepackage{indentfirst}

\begin{document}


\title{
        Pseudospin Symmetry: Microscopic origin of the ground-state inversion in neutron-rich odd-A Cu isotopes\\
       } 

\author{Tianshuai Shang}
\affiliation{College of Physics, Jilin University, Changchun 130012, China}

\author{Qiang Zhao}
\email{qiang.zhao@ibs.re.kr}
\affiliation{Center for Exotic Nuclear Studies, Institute for Basic Science, Daejeon 34126, South Korea}

\author{Jian Li}
\email{jianli@jlu.edu.cn}
\affiliation{College of Physics, Jilin University, Changchun 130012, China}

\date{\today}

\begin{abstract}
The role of the pseudospin symmetry in the inversion of the ground-state spin in Cu isotopes is investigated within the covariant density functional theory (CDFT).
The density functional PC-PK1 gives a good agreement with the observed ground-state properties of the odd-A $^{69-79}$Cu.
In particular, the inversion of the ground-state spin at $A=75$ is reproduced nicely, which can be explained by the level crossing between the proton pseudospin partners $\pi 2p_{3/2}$ and $\pi 1f_{5/2}$.
It is found that the pseudospin symmetry plays a dominant role in the evolution of the energy splittings between the pseudospin partners.
By comparing with the results of the nonrelativistic DFTs, our studies suggest that the pseudospin symmetry is essential to the shell evolution around this region.
\end{abstract}

\pacs{Valid PACS appear here}
\maketitle



The shell evolution towards the extreme neutron-to-proton ratio has been a pivotal focus in nuclear physics over recent decades \cite{sorlin2008_PiPaNP61_602, otsuka2020_RMP92_015002}, since it is crucial to
understand the effective interaction between nucleons and the nucleosynthesis process, particularly the r-process.
Significant efforts have been dedicated to deciphering the mechanism behind the fading of traditional magic numbers and the emergence of new ones.
One of the representatives is the discovery of the doubly magic property of the neutron-rich isotope $^{78}$Ni, which has 28 protons and 50 neutrons \cite{xu2014.PRL113.032505, olivier2017.PRL119.192501, welker2017.PRL119.192502, sahin2017.PRL118.242502, taniuchi2019.N569.53}.

With an additional proton, the Cu isotopes become good candidates for studying the evolution of the single-particle level structure around this region.
Researches have demonstrated the lowering of the first $5/2^-$ state in odd-$A$ $\rm ^{69-73}Cu$ isotopes \cite{franchoo1998PRL81.3100, franchoo2001PRC64.054308} and the change of the ground-state spin-parity from $3/2^-$ in $\rm ^{73}Cu$ to $5/2^-$ in $\rm ^{75}Cu$ \cite{flanagan2009PRL103.142501}.
The two states $3/2^-$ and $5/2^-$ are found to have important single-particle characters \cite{franchoo2001PRC64.054308, stefanescu2008_PRL100_112502, flanagan2009PRL103.142501}, implying a crossing between the single-proton levels $\pi 2p_{3/2}$ and $\pi 1f_{5/2}$.
The shell model calculations show that the lowering of the $5/2^-$ state can be attributed to the tensor force in the monopole interaction \cite{franchoo1998PRL81.3100, otsuka2005.PRL95.232502, otsuka2010.PRL104.012501, sieja2010_PRC81_061303, ichikawa2019.NP15.321}, which is attractive between the proton level $\pi 1f_{5/2}$ and the neutron level $\nu 1g_{9/2}$.
Note that the tensor force would also give a rise of $\pi 1f_{7/2}$ and reduce the $Z=28$ gap, which may lead to a large $\pi 1f_{7/2}$ strength in the low-lying $7/2^-$ states approaching $N=50$ \cite{delafosse2018_PRL121_192502}.
However, this has not been observed in experiment \cite{morfouace2015.PLB751.306,sahin2017.PRL118.242502,olivier2017.PRL119.192501}.
The studies with the nonrelativistic density functional theory (DFT) also show that the inclusion of the tensor force can reproduce the crossing of the $\pi 2p_{3/2}$ and $\pi 1f_{5/2}$ levels in Ni isotopes \cite{brink2018.PRC97.064304, routray2021.PRC104.L011302}.
It is also found that the level crossing can be reproduced without the tensor force by using a simple effective interaction (SEI) \cite{routray2021.PRC104.L011302,bano2022.PRC106.024313}. Therefore, the mechanism behind the level crossing is still unsettled.

The two relevant single-particle levels $\pi 2p_{3/2}$ and $\pi 1f_{5/2}$ are often referred to as the so-called pseudospin doublets that should be degenerate when the pseudospin symmetry is precisely upheld.
The pseudospin symmetry was introduced \cite{arima1969_PLB30_517,hecht1969.NPA137.129} to explain the near degeneracy between the two single-particle states with quantum numbers $(n,l,j=l+1/2)$ and $(n-1,l+2,j=l+3/2)$, which can be defined as the pseudospin doublets with the pseudo-radial, pseudo-orbital, and pseudospin quantum numbers $(\tilde{n}=n-1, \tilde{l}=l+1, \tilde{s}=1/2)$.
The pseudospin symmetry is manifested as a relativistic symmetry of the Dirac Hamiltonian, and $\tilde{l}$ is the orbital angular momentum of the lower component of the Dirac spinor \cite{ginocchio1997.PRL78.436}.
Extensive studies of the pseudospin symmetry have been performed within the relativistic framework,
and an overview can be referred to Refs. \cite{ginocchio2005_PR414_165, liang2015_PR570_1}.
Notice that the pseudospin symmetry can play important roles in the shell structure \cite{long2007.PRC76.034314, long2009.PLB680.428, jolos2007.PAN70.812, li2014_PLB732_169}, and can explain the shape coexistence in the $^{78}$Ni region \cite{delafosse2018_PRL121_192502}.
Thus it is necessary to study the inversion of the pseudospin doublets $\pi 2p_{3/2}$ and $\pi 1f_{5/2}$ in Cu isotopes from the view of the pseudospin symmetry.

The covariant density functional theory (CDFT) is one of the most successful microscopic theories for a global description of nuclei \cite{Meng2016_book}.
It has several advantages due to the inclusion of the Lorentz symmetry \cite{ring2012_PST150_014035}.
For instance, it naturally includes the spin and pseudospin degrees of freedom simultaneously without requiring adjustable parameters, which can explain the large spin-orbit potential in nuclei and the origin of the pseudospin symmetry.
Moreover, the theory adeptly handles time-odd fields, crucial for detailing spectroscopic features linked to nuclear rotations \cite{afanasjev2010_PRC82_034329, meng2013_FP8_55} and their nuclear magnetic moments \cite{furnstahl1989.PRC40.1398, hofmann1988.PLB214.307, yao2006.PRC74.024307}.
CDFT's applications have consistently proven their mettle in analyzing both static and dynamic nuclear properties\cite{meng2006_PiPaNP57_470, niksic2011_PiPaNP66_519, ring1996_PiPaNP37_193, vretenar2005_PR409_101, ren2020_PLB801_135194, ren2022_PRC105_L011301, ren2022_PRL128_172501}.

One of the widely used CDFTs is the relativistic point-coupling model \cite{burvenich2004_NPA744_92, nikolaus1992_PRC46_1757, zhao2010.PRC82.054319}, where the energy density functional is built upon contact interactions.
It is much simpler in numerical applications with both the mean-field theory and the beyond mean-field theory compared to the meson-exchange CDFT.
This model not only paves the way for a deeper exploration of its ties to nonrelativistic approaches \cite{ren2020_PRC102_021301, sulaksono2003_AoP308_354} but also facilitates the effortless inclusion of exchange terms via the Fierz transformation \cite{sulaksono2003_AoP306_36, liang2012_PRC86_021302R, zhao2022.PRC106.034315}.
Due to these advantages, the relativistic point-coupling model has attracted a lot of attention.

In this letter, we investigate the inversion of the ground-state spin in Cu isotopes using the CDFT framework that naturally includes the pseudospin degree of freedom.
The point-coupling covariant density functional PC-PK1 \cite{zhao2010.PRC82.054319} is employed.
It has been successfully applied to describe the ground-state properties \cite{xia2018_ADNDT121-122_1, yang2021_PRC104_054312, zhang2022_ADaNDT144_101488}, the collective excitations \cite{li2012_PLB717_470, zhao2011_PRL107_122501, zhao2017_PLB773_1,zhao2016Phys.Rev.C94.041301, wang2022_PRC105_054311},
and the dynamic properties \cite{ren2022_PRC105_L011301, ren2022_PRL128_172501}.
In particular, it provides an accurate description on the isospin dependence of the binding energy \cite{zhao2010.PRC82.054319, zhao2012Phys.Rev.C86.064324, zhang2021_PRC104_L021301}.
Notably, we find that PC-PK1 can reproduce the inversion of the ground-state spin in $^{75}$Cu without including the tensor force explicitly. This sheds light on the potential influence of pseudospin symmetry in this observed phenomenon.


A detailed theoretical framework of the relativistic point-coupling model can be found in Refs. \cite{burvenich2004_NPA744_92, nikolaus1992_PRC46_1757, zhao2010.PRC82.054319}.
For a quantitative analysis of the pseudospin symmetry's effects, we need to derive the contribution of the pseudospin-orbit potentials to the single-particle energy.
Assuming the spherical symmetry, the general radial Dirac equation for spherical nuclei reads
\begin{equation}
  \label{DEQ}
  \begin{pmatrix}
  	V+S & -\frac{d}{dr} + \frac{\kappa}{r} \\
  	\frac{d}{dr} + \frac{\kappa}{r} & V-S-2M
  \end{pmatrix}
  \begin{pmatrix}
  	G_a \\
  	F_a
  \end{pmatrix}
  =
  \varepsilon_a
  \begin{pmatrix}
  	G_a \\
  	F_a
  \end{pmatrix},		
\end{equation}
in which $\kappa=(l-j)(2j+1)$, $M$ is the nucleon mass, $V$ is the vector potential, $S$ is the scalar potential, $\varepsilon_a$ is the single-particle energy of the level $a$, $G_a$ and $F_a$ correspond to the upper and lower part of the Dirac spinor, respectively.
From Eq.~(\ref{DEQ}), a Schr\"odinger-like equation can be derived for the lower component $F$ as
\begin{align}\label{equ:schrodinger}
    \bigg[ \frac{1}{M_-}\frac{d^2}{dr^2}
    & + \frac{1}{M_-}\frac{\kappa(\kappa-1)}{r^2} + \Sigma_+
      + \frac{1}{M_-M_+}\frac{d\Sigma_+}{dr}\frac{\kappa}{r} \nonumber \\
    & \qquad - \frac{1}{M_-M_+}\frac{d\Sigma_+}{dr}\frac{d}{dr} \bigg] F_a = \varepsilon_a F_a,
\end{align}
where $\Sigma_+ = V + S$, $M_+ = \Sigma_+ - \varepsilon_a$, $\Sigma_- = V - S - 2M$, and $M_- = \Sigma_- - \varepsilon_a$.
The above Schr\"odinger-like equation contains
the pseudospin-orbit potential $V_{\rm PSO}$
\begin{align}
  V_{\rm PSO} = \frac{1}{M_-M_+}\frac{d\Sigma_+}{dr}\frac{\kappa}{r}.
\end{align}
The exact pesudospin symmetry requires that $\Sigma_{+}=0$ or $d\Sigma_{+}/dr=0$ \cite{ginocchio1997.PRL78.436, meng1998_PRC58_R628}.
However, both conditions give no bound states, and thus the pseudospin symmetry is not exactly conserved in realistic nuclei.

The contribution of the pseudospin-orbit potential $V_{\rm PSO}$ to the single-particle energy is obtained by taking its expectation on the lower component of the Dirac spinor as
\begin{equation}\label{equ:expectation}
  \varepsilon_a^{\rm PSO}
  = \frac{1}{\int dr~ F_a^2}~ \mathcal{P}\int dr~ \frac{1}{M_-M_+}\frac{d\Sigma_+}{dr}\frac{\kappa}{r}F_a^2,
\end{equation}
where $\mathcal{P}$ denotes the principal value of the integral.
Similarly, the contributions of other terms in the left-hand side (l.h.s) of Eq. (\ref{equ:schrodinger}) can be calculated.


In order to address both the mean field and the pairing field within a consistent framework, the relativistic Hartree-Bogoliubov (RHB) method is utilized.
The RHB equations are solved in the space of the spherical Harmonic oscillator basis \cite{niksic2014_CPC185_1808} with 30 major shells.
A separable pairing force is taken in the particle-particle channel with the two parameters $G=728\text{~MeV}\cdot\text{fm}^3$ and $a=0.644\text{~fm}$ \cite{tian2009.PLB676.44}.
Given that Cu isotopes possess an odd number of protons, the blocking effect is considered for the unpaired proton.
Calculations with the blocked proton levels around Fermi energy ($\pi 1f_{5/2}$, $\pi 2p_{3/2}$ and $\pi 2p_{1/2}$) are performed.
The configuration with the lowest energy is determined as the ground state, while others are approximated as the excited states.
The total energy difference between the ground state and the excited state is estimated as the excited energy.

\begin{table}[htbp]
  \centering
  \caption{The ground-state spin-parity and the binding energy $E_B$ (in the unit of MeV)  for Cu isotopes are computed using the covariant density functional PC-PK1 \cite{zhao2010.PRC82.054319}, DD-ME2 \cite{lalazissis2005.PRC71.024312}, and the simple effective interaction (SEI) \cite{routray2021.PRC104.L011302} in comparison with the experimental data \cite{wang2021_CPC45_030003}. The root-mean-square (\emph{rms}) deviations of the theoretical binding energies to the experimental data are shown in the last row.}
  \label{tab:ground-state}
  \begin{tabular}{cccccccccccc}
  \hline\hline
  \multirow{2}{*}{A} & \multicolumn{2}{c}{EXP} & & \multicolumn{2}{c}{PC-PK1} & & \multicolumn{2}{c}{DD-ME2} & & \multicolumn{2}{c}{SEI} \\ \cline{2-3} \cline{5-6} \cline{8-9} \cline{11-12}
     & $J^{\pi}$ & $E_B$ & & $J^{\pi}$ & $E_B$ & & $J^{\pi}$ & $E_B$ & & $J^{\pi}$ & $E_B$ \\ \hline
  69 & $3/2^-$ &  599.97 & & $3/2^-$ &  600.23 & & $3/2^-$ &  599.32 & & $3/2^-$ &  598.59 \\
  71 & $3/2^-$ &  613.09 & & $3/2^-$ &  613.06 & & $5/2^-$ &  611.38 & & $3/2^-$ &  612.93 \\
  73 & $3/2^-$ &  625.51 & & $3/2^-$ &  624.85 & & $5/2^-$ &  622.97 & & $3/2^-$ &  625.76 \\
  75 & $5/2^-$ &  637.13 & & $5/2^-$ &  636.30 & & $5/2^-$ &  634.02 & & $3/2^-$ &  637.49 \\
  77 & $5/2^-$ &  647.67 & & $5/2^-$ &  647.06 & & $5/2^-$ &  644.61 & & $5/2^-$ &  648.38 \\
  79 & $5/2^-$ &  657.35 & & $5/2^-$ &  657.05 & & $5/2^-$ &  654.75 & & $5/2^-$ &  658.19 \\
  \emph{rms} &         &         & &         &    0.52 & &         &    2.43 & &         &    0.74 \\
  \hline\hline
  \end{tabular}
\end{table}

\begin{figure}[htbp]
\centering
\includegraphics[width=0.45\textwidth]{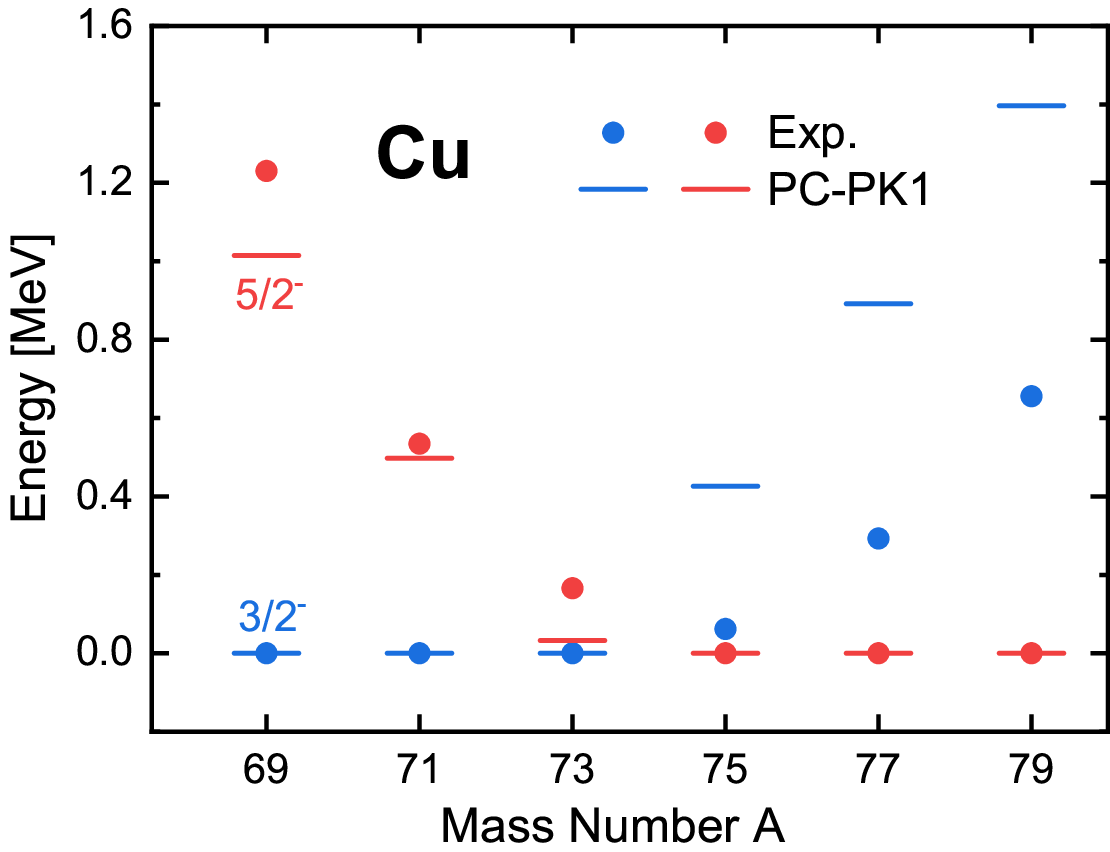}
\caption{(Color online) Systematics of the first $3/2^-$ and $5/2^-$ states in odd-$A$ $\rm ^{69-79}Cu$ isotopes calculated with functional PC-PK1~(lines) in comparison with the experimental data~(dots)~\cite{zeidman1978.PRC18.2122, franchoo1998PRL81.3100, sahin2017.PRL118.242502, olivier2017.PRL119.192501}.}
\label{energy}
\end{figure}

The ground-state spin-parities and the total binding energies for Cu isotopes given by the density functional PC-PK1 are listed in Table \ref{tab:ground-state}.
Results obtained with the finite-range covariant density functional DD-ME2 \cite{lalazissis2005.PRC71.024312} and the simple effective interaction (SEI) \cite{routray2021.PRC104.L011302} are also shown.
While all the theoretical calculations anticipate the inversion of the ground-state spin, only the results of PC-PK1 agree with the observations.
Compared to DD-ME2 and SEI, PC-PK1 offers a more accurate depiction of the binding energies with the root-mean-square (\emph{rms}) deviations approximating 0.52 MeV.
The calculated excited energies are depicted in Fig. \ref{energy} in comparison with the experimental values.
It can be seen that the results of PC-PK1 align well with the observed evolution tendency, especially for the rapid lowering of the $5/2^-$ state and the inversion of the ground-state spin at $^{75}$Cu.

\begin{figure}[htbp]
\centering
\includegraphics[width=0.45\textwidth]{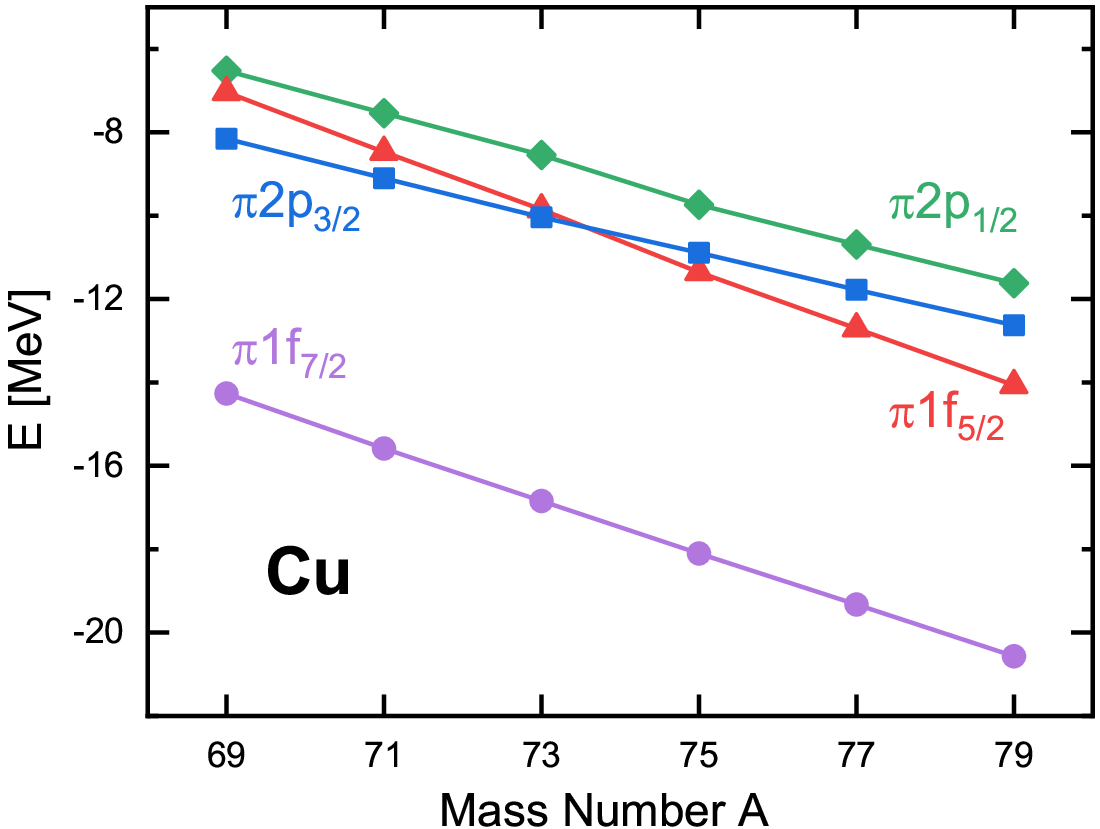}
\caption{(Color online) Proton single-particle levels of $\pi 1f_{7/2}$, $\pi 2p_{3/2}$, $\pi 1f_{5/2}$ and $\pi 2p_{1/2}$ for odd-$A$ $\rm ^{69-79}Cu$ isotopes calculated with the density functional PC-PK1.}
\label{level}		
\end{figure}

The inversion of the ground-state spin can be understood via the proton single-particle levels in Fig.~\ref{level}.
Since all the levels below the shell gap $Z=28$ are fully occupied, the ground-state spin would be determined by the level occupied by the last proton with the lowest energy.
Due to the level crossing between $\pi 2p_{3/2}$ and $\pi 1f_{5/2}$, the ground state is changed from $3/2^-$ at $^{73}$Cu to $5/2^-$ at $^{75}$Cu.
Many studies based on the shell models and the nonrelativistic DFTs show that the tensor force plays an important role in explaining the level crossing.
But the level crossing also corresponds to the restoration and breaking of the pseudospin symmetry, where the pseudospin degree of freedom should be important.
Thus it is prior to study this phenomena under the framework of the CDFT.

\begin{figure}[htbp]
\centering
\includegraphics[width=0.45\textwidth]{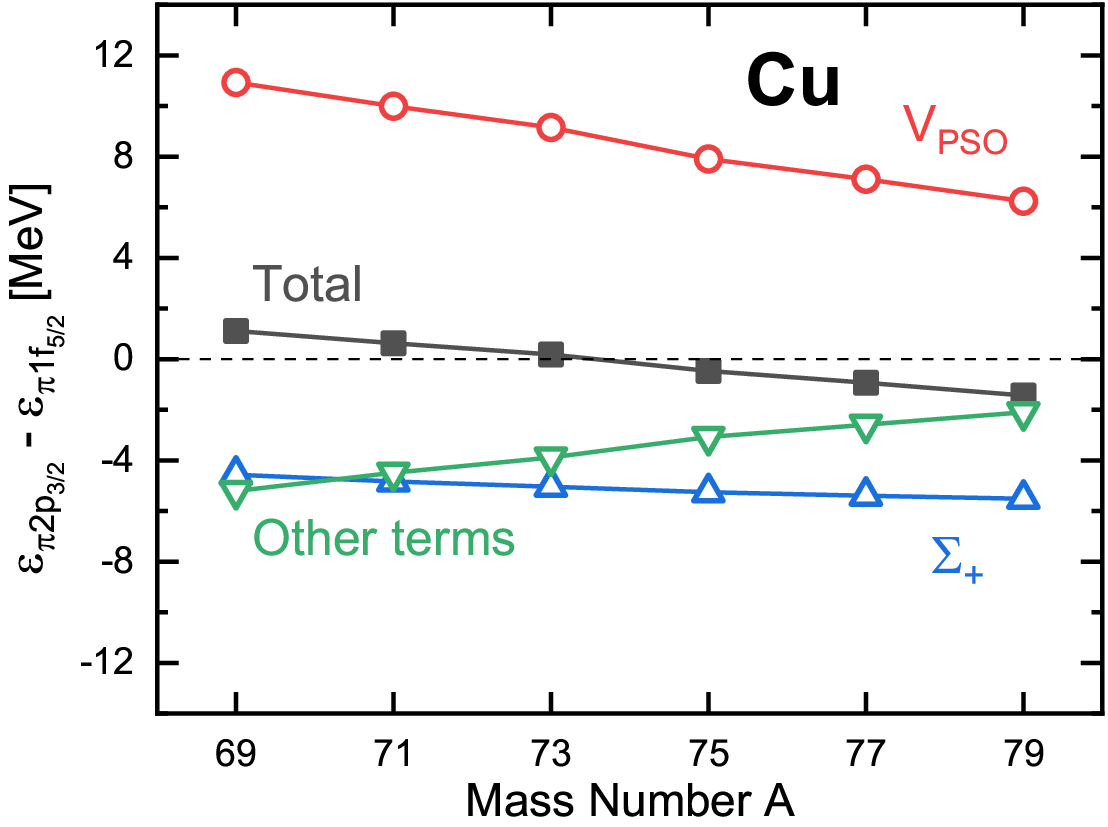}
\caption{(Color online) The pseudospin-orbit splittings between proton levels $\pi 2p_{3/2}$ and $\pi 1f_{5/2}$ for odd-$A$ $\rm ^{69-79}Cu$ isotopes calculated with the density functional PC-PK1.
The contributions from $V_{\rm PSO}$, $\Sigma_+$, and the other terms in Eq. (\ref{equ:schrodinger}) are also shown.}
\label{PSS}
\end{figure}

Figure \ref{PSS} presents the energy splittings between the pseudospin partners ($\pi 2p_{3/2}$, $\pi 1f_{5/2}$) calculated by the density functional PC-PK1.
The contributions of the different terms in the Schr\"odinger-like Eq. (\ref{equ:schrodinger}) can be obtained by using Eq. (\ref{equ:expectation}), where $F_a$ is taken the lower component of the canonical wave function in the RHB calculations.
The total energy splittings are less than 1.5 MeV and, in particular, the pseudospin symmetry is nicely restored at $^{73}$Cu.
Several characters can be seen in Fig. \ref{PSS}.
1) The contributions of the pseudospin-orbit potential are much larger than the total energy splittings.
2) The contributions of other terms in Eq. (\ref{equ:schrodinger}) are also larger than the total energy splittings and comparable to the contributions of the pseudospin-orbit potential.
3) The total energy splittings are inverted after $^{73}$Cu, but the contributions from the pseudospin-orbit potential do not change the sign.
All these findings infer that the pseudospin-orbit potential can not be treated as a perturbative quantity \cite{marcos2001_PLB513_30, alberto2002_PRC65_034307}.
The quasi-degeneracy of the pseudospin partners is a result of the cancellation between the pseudospin-orbit potentials and other terms in Eq. (\ref{equ:schrodinger}).
The pseudospin symmetry is also interpreted as a kind of dynamical symmetry \cite{alberto2001.PRL86.5015, arima1969_PLB30_517}.

It is important to study the evolution of the two pseudospin partners $\pi 2p_{3/2}$ and $\pi 1f_{5/2}$ with the increasing mass number.
Note that the energy splittings between the two levels at $^{69}$Cu is about 1.11 MeV, which is comparable with the experimental excited energy 1.23 MeV for the $5/2^-$ state.
With the increasing mass number, the total energy splittings decrease continuously to -1.44 MeV.
The reduction of the energy splittings is mainly dominated by the pseudospin-orbit potential $V_{\rm PSO}$ and the mean-field potential $\Sigma_+$, while other terms together give an opposite effect as seen in Fig. \ref{PSS}.
Compared to $\Sigma_+$ which reduces the energy splittings by 0.96 MeV, $V_{\rm PSO}$ gives more contributions with a value of 4.69 MeV.
It can be seen that $V_{\rm PSO}$ plays a major role in the evolution of $\pi 2p_{3/2}$ and $\pi 1f_{5/2}$.

\begin{figure}[htbp]
\centering
\includegraphics[width=0.45\textwidth]{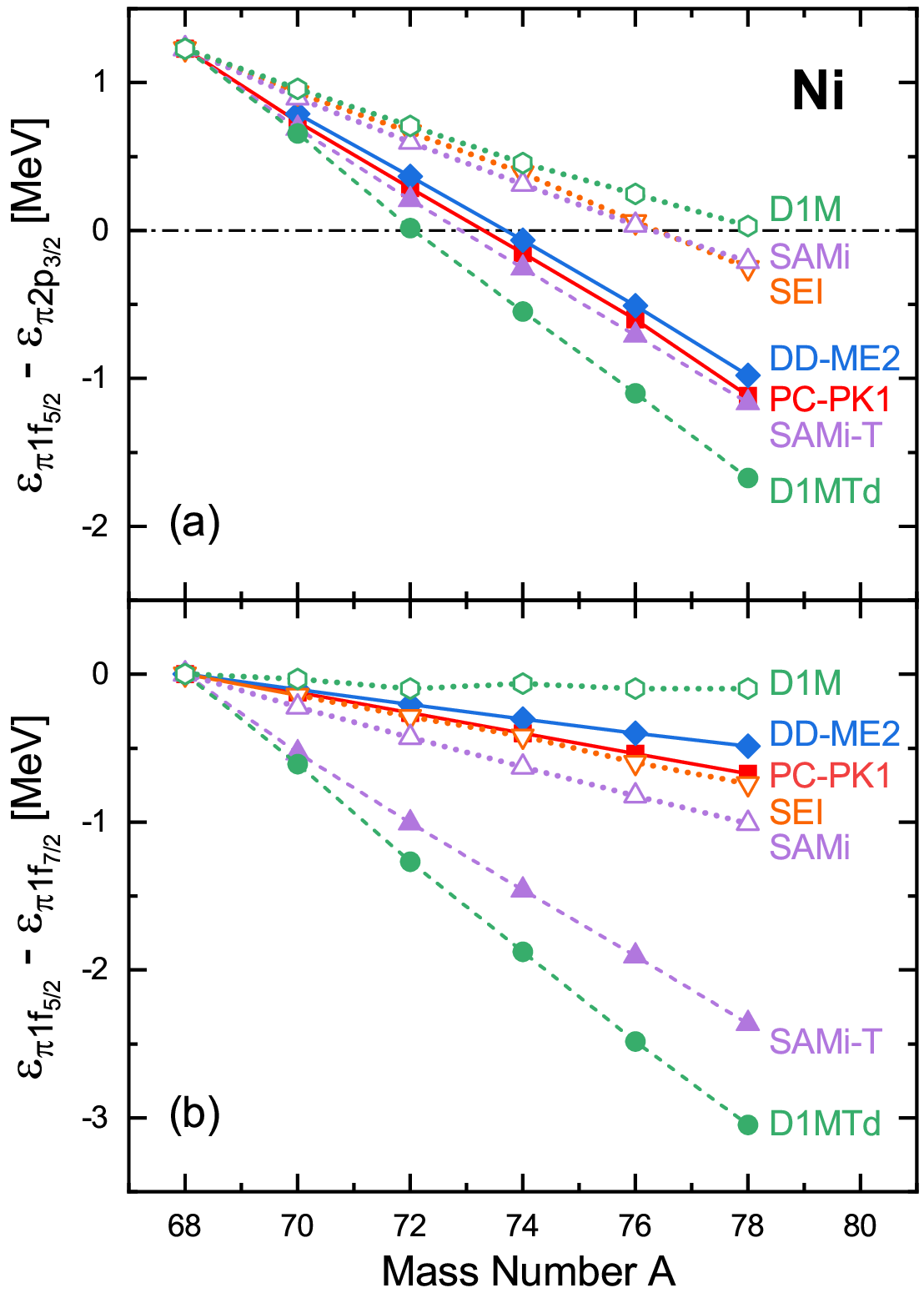}
\caption{(Color online) Energy splittings of the pseudospin partners $(\pi 2p_{3/2},\pi 1f_{5/2})$ (a) and the spin partners ($\pi 1f_{5/2}, \pi 1f_{7/2}$) (b) given by the covariant density functional PC-PK1 and DD-ME2, in comparison with the nonrelativistic calculations using the Skyrme interactions (SAMi and SAMi-T), the Gogny interaction (D1M and D1MTd), and the simple effective interaction (SEI). All the results are taken $A=68$ as a reference. The pseudospin-orbit splittings are shifted to meet the experimental energy difference between $5/2^-$ and $3/2^-$ states in $^{69}$Cu. }
\label{gap}
\end{figure}

Since both the pseudospin symmetry and the tensor force can have important influences on shell evolution, it is better to further illustrate the differences between these two mechanisms.
Figure \ref{gap} (a) shows the evolutions of the pseudospin-orbit splittings between $\pi 2p_{3/2}$ and $\pi 1f_{5/2}$ given by the nonrelativistic models and relativistic models for Ni isotopes.
The evolution of the single-particle levels in Ni isotopes is similar to the results in Cu isotopes, which would not change our conclusions here.
For the nonrelativistic models, we employ the effective interactions without the tensor force (the simple effective interaction SEI \cite{routray2021.PRC104.L011302}, the Skyrme interaction SAMi\cite{roca-maza2012Phys.Rev.C86.031306}, and the Gongy interaction D1M \cite{goriely2009_PRL102_242501,routray2021.PRC104.L011302}) and the ones with the tensor force (SAMi-T \cite{shen2019Phys.Rev.C99.034322} and D1MTd \cite{co2018_PRC97_034313,routray2021.PRC104.L011302}).
The calculated results are shifted to meet the experimental data 1.23 MeV, which is taken as the energy difference between $5/2^-$ and $3/2^-$ states in $^{69}$Cu.
It is seen that, for the nonrelativistic models without tensor force, the evolutions of the energy splittings are too weak to present the level crossing before $^{76}$Ni.
After including the tensor force, the evolutions of the energy splittings are enhanced and the level crossings occur between $^{76}$Ni and $^{78}$Ni.
This is due to the tensor-force effect, which is attractive between $\pi 1f_{5/2}$ and $\nu 1g_{9/2}$, and repulsive between $\pi 2p_{3/2}$ and $\nu 1g_{9/2}$.
However, even though the tensor force is not taken into account explicitly, the covariant density functionals can give strong evolutions similar to the results of the nonrelativistic calculations with the tensor force.
This illustrates that the influence of the pseudospin symmetry is comparable to that of the tensor force.
Thus the pseudospin mechanism should be taken into account when we study the inversion of the ground-state spin in Cu isotopes.

The situation can be different for the spin-orbit splittings. In Fig. \ref{gap} (b), the spin-orbit splittings of $\pi 1f$ orbitals given by the covariant DFTs and the nonrelativistic models are presented.
In contrast to the case of the pseudospin-orbit splittings, the relativistic models present weak evolutions of the spin-orbit splittings as the nonrelativistic models without the tensor force.
This indicates the pseudospin symmetry's nature that it mainly influences the pseudospin partners rather than the spin partners.
Note that the large reduction of the spin-orbit splittings of $\pi 1f$ may lead to a large $\pi f_{7/2}$ strength in the low-lying $7/2^-$ states, which has not been observed so far \cite{morfouace2015.PLB751.306, sahin2017.PRL118.242502, olivier2017.PRL119.192501}.
In contrast to the tensor-force effect, the pseudospin symmetry provides a nice explanation of this feature.
Further precise experiments in the future can be helpful in shedding light on this subject.


In summary, the inversion of the ground-state spin in Cu isotopes has been studied within the covariant density functional theory.
Different from the nonrelativistic models, the covariant density functional PC-PK1 can reproduce the ground-state inversion at $^{75}$Cu without the tensor force.
It is found that the pseudospin symmetry plays an important role in the level crossing between the proton pseudospin partners $\pi 2p_{3/2}$ and $\pi 1f_{5/2}$ leading to the ground-state inversion.
The effect of the pseudospin symmetry on the evolution of the pseudospin-orbit splittings can be comparable to the tensor force effect in the nonrelativistic DFTs.
It suggests that the pseudospin symmetry can be important in this region and should be taken into account.
Further studies by taking into account beyond mean-field effects and combining the future experimental observations are expected.

\section*{Acknowledgments}
The authors thank P.W. Zhao for helpful discussions and Z.Z. Li for providing calculations with the Skyrme interactions.
This work is supported by the National Natural Science Foundation of China (No 11675063), the Natural Science Foundation of Jilin Province (No. 20220101017JC), and the Key Laboratory of Nuclear Data Foundation (JCKY2020201C157) as well as the Institute for Basic Science funded by the Korean government (Grant No. IBS-R031-D1).

\bibliography{reference}

\end{document}